\documentclass[onecolumn]{article}
\usepackage{amsfonts}
\usepackage{amsmath}
\usepackage{graphicx}
\usepackage[dvipdf]{color}

\input{tcilatex}

\begin{document}

\title{Nondistillability of distillable qutrit-qutrit states under
depolarizing noise}
\author{Salman Khan\thanks{%
sksafi@phys.qau.edu.pk}, M. K. Khan \\
Department of Physics, Quaid-i-Azam University, \\
Islamabad 45320, Pakistan}
\maketitle

\begin{abstract}
We study the effects of decoherence on some particular bipartite qutrit
states under the influence of global, collective, local and multilocal
depolarizing noise. We show that certain free entangled distillable qutrit
density matrices become bound entangled or separable and hence convert into
nondistillable density matrices in global noise. The collective noise
increases the degree of entanglement of the qutrit bipartite states.
Furthermore, we show that some particular local operation cannot avoid the
Nondistillability of the distillable states.\newline

\end{abstract}

\section{Introduction}

Entanglement is one of the potential sources of quantum theory. It is the
key concept and a major resource for quantum communication and computation 
\cite{springer}. In the last two decades, enormous efforts have been made to
investigate various aspects of quantum entanglement and its benefits in a
number of setups, such as teleportation of unknown states \cite{Bennett1},
quantum key distribution \cite{Ekert}, quantum cryptography \cite{Bennett2}
and quantum computation \cite{Grover, Vincenzo}. However, a realistic
quantum system cannot be completely isolated from its environment.
Generally, the coupling between an environment and a quantum system leads to
the phenomenon of decoherence and it gives rise to an irreversible transfer
of information from the system to the environment \cite{Zurik, Breuer,
Zurik2}. Nevertheless, It has also been shown \cite{Braun, Benatti,Nayak}
that the environment can induced entanglement between two otherwise totally
decoupled systems. It is, thus, necessary to investigate the behavior of
initial state entanglement in the presence of decoherence. Yu and Eberly 
\cite{Yu1, Yu2} showed that entanglement loss occurs in a finite time under
the action of pure vacuum noise in a bipartite state of qubits. They found
that, even though it takes infinite time to complete decoherence locally,
the entanglement may be lost in finite time. The name "entanglement sudden
death" (ESD) was tossed for the phenomenon of sudden loss of entanglement.
The finite time loss of entanglement definitely limits the application of
entangled states in quantum information processing. The phenomenon of ESD is
not limited only to two qubit entangled state, it is investigated in systems
of larger spaces such as qutrits and qudits \cite{Yonac, Jakobczyk, Ikram,
Qasimi, Jeager1, Jeager2, Haung, Jeager3}. A geometric interpretation of the
effect of ESD is given in Ref. \cite{Terra}. The experimental evidences of
the phenomenon of ESD have been reported for optical setups \cite{Almeida}
and atomic ensembles \cite{Laurat}.

The extraction of a pure entangled singlet state from an ensemble of mixed
state through local quantum operation and classical communication (LOCC) is
called distillation \cite{Bennett3}. An inseparable state is said to possess
some degree of entanglement. Quantum states are grouped into separable and
entangled states for qubit-qubit and qubit-qutrit system by using
Peres-Horodecki criterion \cite{Peres, Horodecki1}. According to this
criterion, the partial transpose of a separable density matrix must has
non-negative eigenvalues. The partial transpose of a bipartite density
matrix $\rho _{m\nu ,n\mu }$ over the second qutrit B is given by $\rho
_{m\mu ,n\nu }^{T_{B}}=\rho _{m\nu ,n\mu }$ and for the first qutrit, it can
similarly be defined. Nevertheless, such a characterization for higher
dimensional bipartite states is difficult \cite{Horodecki2}. A bipartite
entangled state can either be free or bound entangled state. A free
entangled state can be distilled to a singlet state through LOCC whereas a
bound entangled state cannot be, no matter how many copies are available. A
bound entangled state cannot be used for reliable quantum information
processing \cite{Horodecki3}. A nondistillable bound entangled state have
positive partial transpose (PPT), whereas a negative partial transpose (NPT)
states are regarded distillable \cite{Horodecki3}. However, no single
criterion is sufficient to detect bound entangled states. For example, for a
qutrit-qutrit system there are many bound entangled states and no single
criterion can fully describe all of them \cite{Clarisse}. However, it is
shown in Ref. \cite{Chen} that realignment criterion can be used to detect
certain bound entangled states. For a bipartite density matrix $\rho _{m\nu
,n\mu }$, the realignment criterion is given by $\left( \rho ^{R}\right)
_{mn,\nu \mu }=\rho _{m\nu ,n\mu }$. A state is separable under realignment
criterion if $\left\Vert \rho ^{R}\right\Vert \leq 1$ and a PPT state is
bound entangled if the quantity $\left\Vert \rho ^{R}\right\Vert -1$ is
positive. This is important to point out here that realignment criterion is
not capable to detect all bound entangled states. The phenomenon in which a
free entangled state becomes nondistillable is called distillability sudden
death (DSD). It is shown in Refs. \cite{Jeager3, Song, Mazher2} that in the
presence of dephasing noise some free entangled states of qutrit-qutrit
system become nondistillable in a finite time. The effect of amplitude
damping channel on such a system is studied in Ref. \cite{Mazher1}, where
the authors found that a simple local unitary transformation can completely
avoid DSD.

In this paper we study the behavior of entanglement by revisiting a
particular family of density matrices that are considered in Refs. \cite%
{Jeager3, Song, Mazher2,Mazher1} of qutrit-qutrit system in the presence of
depolarizing noise. We use partial transpose criterion and realignment
criterion for our investigation. We consider various coupling of the system
and environment in which the system is influenced by global, collective,
local or multilocal depolarizing noise. It is shown that under certain
particular conditions, the depolarizing noise can increase the degree of
entanglement in the qutrit-qutrit system. For example, under the action of
multilocal depolarizing noise, the negativity increases linearly with the
increase of depolarizing noise of one qutrit's local environment by
controlling the other qutrit's local environment. On the other hand, the
free entangled density matrices can become bound entangled or separable and
hence nondistillable under another particular condition in the depolarizing
noise. For example, all the free entangled density matrices become bound
entangled under the action of global environment. Furthermore, it is shown
that DSD cannot be avoided by performing simple local unitary operation,
considered in this paper, for the density matrices. Since no definitive
criterion for separability or entanglement of density matrices of dimensions
greater than six is available, we believe that our results are valid for the
family of density matrices considered in this paper.

\subsection{Qutrit-Qutrit System in a Depolarizing Noise}

We consider a composite system of two qutrits A and B that are coupled to a
noisy environment both collectively and individually. The qutrits are
spatially separated and has no direct interaction with each other. The
collective coupling refers to the situation when both the qutrits are
influenced by the same environment and the multilocal coupling describes the
situation when each qutrit is independently influenced by its own
environment. The system is said to be coupled to a global environment when
it is influenced by both collective and multilocal noises at the same time.
Let the bases of Hilbert space of each qutrit be denoted by $|0\rangle $, $%
|1\rangle $ and $|2\rangle $. Then the bases of the composite system are
given in the order $|00\rangle $, $|01\rangle $, $|02\rangle $, $|10\rangle $%
, $|11\rangle $, $|12\rangle $, $|20\rangle $, $|21\rangle $, $|22\rangle $.

The dynamics of the composite system in the presence of depolarizing noise
can best be described in the Kraus operators formalism. The Kraus operators
for a single qutrit depolarizing noise, which satisfy the completeness
relation $\sum_{i}E_{i}^{\dag }E_{i}=I$, are given as \cite{Salimi}%
\begin{eqnarray}
E_{0} &=&\sqrt{1-p}I_{3},\quad E_{1}=\sqrt{\frac{p}{8}}Y,\quad E_{2}=\sqrt{%
\frac{p}{8}}Z,  \notag \\
E_{3} &=&\sqrt{\frac{p}{8}}Y^{2},\quad E_{4}=\sqrt{\frac{p}{8}}YZ,\quad
E_{5}=\sqrt{\frac{p}{8}}Y^{2}Z,  \notag \\
E_{6} &=&\sqrt{\frac{p}{8}}YZ^{2},\quad E_{7}=\sqrt{\frac{p}{8}}%
Y^{2}Z^{2},\quad E_{8}=\sqrt{\frac{p}{8}}Z^{2},  \label{1}
\end{eqnarray}%
with

\begin{eqnarray}
Y &=&\left( 
\begin{array}{ccc}
0 & 1 & 0 \\ 
0 & 0 & 1 \\ 
1 & 0 & 0%
\end{array}%
\right) ,\quad Z=\left( 
\begin{array}{ccc}
1 & 0 & 0 \\ 
0 & \omega & 0 \\ 
0 & 0 & \omega ^{2}%
\end{array}%
\right) ,  \label{2} \\
I_{3} &=&\left( 
\begin{array}{ccc}
1 & 0 & 0 \\ 
0 & 1 & 0 \\ 
0 & 0 & 1%
\end{array}%
\right)
\end{eqnarray}%
where $\omega =e^{2i\pi /3}$ and $p=1-e^{-\Gamma t/2}\in \left[ 0,1\right] $
is the decoherence parameter. The lower and upper limits of $p$ stand,
respectively, for undecohered and fully decohered cases of the noisy
environment. The evolution of the density matrix when it is influenced by
the global depolarizing noise is given by%
\begin{equation}
\rho (t)=\sum_{i=0}^{80}\sum_{j=0}^{8}\left(
E_{i}^{AB}E_{j}^{B}E_{j}^{A}\right) \rho \left( E_{j}^{A\dag }E_{j}^{B\dag
}E_{i}^{AB\dag }\right) ,  \label{3}
\end{equation}%
where $E_{j}^{A}=E_{m}\otimes I_{3}$, $E_{j}^{B}=I_{3}\otimes E_{m}$ are the
Kraus operators of the multilocal coupling of each individual qutrit and $%
E_{i}^{AB}$ are the Kraus operators of the collective coupling that are
formed from all the possible combinations of the tensor product of the Kraus
operators of a single qutrit depolarizing noise in the form $E_{m}\otimes
E_{m}$. The subscripts $m=0,1,2,...8$ stand for a single qutrit Kraus
operators of a depolarizing noise given in Eq. (\ref{1}). The system is said
to be under the action of global depolarizing noise when both multilocal and
collective couplings are switched on simultaneously (global = multilocal +
collective). We consider the following initial density matrix of the
bipartite qutrit system%
\begin{equation}
\rho (0)=\frac{2}{7}|\Psi _{+}\rangle \langle \Psi _{+}|+\frac{\alpha }{7}%
\sigma _{+}+\frac{5-\alpha }{7}\sigma _{-},  \label{4}
\end{equation}%
where $2\leq \alpha \leq 5$. In Eq. (\ref{4}) $|\Psi _{+}\rangle $ is a
maximally entangled bipartite qutrit state given by $|\Psi _{+}\rangle =1/%
\sqrt{3}\left( |00\rangle +|11\rangle +|22\rangle \right) $, $\sigma _{+}$
and $\sigma _{-}$ are separable states, which are given by $\sigma
_{+}=1/3\left( |01\rangle \langle 01|+|12\rangle \langle 12|+|20\rangle
\langle 20|\right) $, $\sigma _{-}=1/3\left( |10\rangle \langle
10|+|21\rangle \langle 21|+|02\rangle \langle 02|\right) $. The density
matrix of Eq. (\ref{4}) is separable for $2\leq \alpha \leq 3$, bound
entangled for $3\leq \alpha \leq 4$ and free entangled for $4<\alpha \leq 5$ 
\cite{Horodecki4}. For a further detailed study of the properties of these
density matrices we refer the readers to Refs. \cite{Clarisse,Chen,Chen2}.
We will concentrate on the entanglement behavior of free entangled density
matrix in the range $4<\alpha \leq 5$. Using initial density matrix of Eq. (%
\ref{4}) in Eq. (\ref{3}) and taking the partial transpose over the second
qutrit, it is easy and straightforward to find the eigenvalues. Let the
decoherence parameters for multilocal noise of the two qutrits and
collective noise of the composite system be $p_{1}$, $p_{2}$ and $p$
respectively. Then, the eigenvalues of the partial transpose of the final
density matrix of Eq. (\ref{3}) when only the first qutrit is coupled to the
noisy environment are given by%
\begin{eqnarray}
\lambda _{1,3,5} &=&\frac{1}{336}\left( 40-3p_{1}+\sqrt{\left( 41-20\alpha
+41\alpha ^{2}\right) \left( 9p_{1}-8\right) ^{2}}\right)  \notag \\
\lambda _{2,4,6} &=&\frac{1}{336}\left( 40-3p_{1}-\sqrt{\left( 41-20\alpha
+41\alpha ^{2}\right) \left( 9p_{1}-8\right) ^{2}}\right)  \notag \\
\lambda _{7,8,9} &=&\frac{2}{21}+\frac{1}{56}p_{1}  \label{5}
\end{eqnarray}%
The eigenvalues of the partial transpose of the final density matrix when
both the qutrits are coupled to multilocal depolarizing noise are given by%
\begin{eqnarray}
\lambda _{1,3,5} &=&\frac{1}{2688}(320-24p_{2}+3p_{1}(-8+9p_{2})  \notag \\
&&+\sqrt{(41-20\alpha +4\alpha ^{2})(-8+9p1)^{2}(-8+9p2)^{2}})  \notag \\
\lambda _{2,4,6} &=&\frac{1}{2688}(320-24p_{2}+3p_{1}(-8+9p_{2})  \notag \\
&&-\sqrt{(41-20\alpha +4\alpha ^{2})(-8+9p_{1})^{2}(-8+9p_{2})^{2}})  \notag
\\
\lambda _{7,8,9} &=&\frac{1}{1344}(128+24p_{1}+24p_{2}-27p_{1}p_{2})
\label{6}
\end{eqnarray}%
It is easy to see that Eq. (\ref{6}) reduces to Eq. (\ref{5}), if we switch
off the coupling of second qutrit with its local environment. The
eigenvalues when both qutrits are coupled to the same environment, that is,
for collective coupling of the environment and system are obtained by
replacing $p_{1}=p_{2}=p$ in Eq. (\ref{6}). The only eigenvalues that
possibly become negative in Eqs. (\ref{5}), (\ref{6}) and hence for the
collective noise are those which are given by $\lambda _{2,4,6}$. To observe
the behavior of entanglement and see whether the phenomena of DSD and ESD
happen in these cases, we use the partial transpose criterion and
realignment criterion in the following.

The degree of entanglement in NPT states is quantified by using negativity 
\cite{Peres, Horodecki1}. It is given by the sum of the absolute value of
the negative eigenvalues of the partial transpose of a density matrix. The
negativity for local, multilocal and collective noise becomes%
\begin{equation}
\mathcal{N}(\rho )=\max \left\{ 0,\left\vert \lambda _{2,4,6}\right\vert
\right\} =\max \left\{ 0,\left\vert 3\lambda _{2}\right\vert \right\} .
\label{7}
\end{equation}%
\begin{figure}[h]
\begin{center}
\begin{tabular}{ccc}
\vspace{-0.5cm} \includegraphics[scale=1.2]{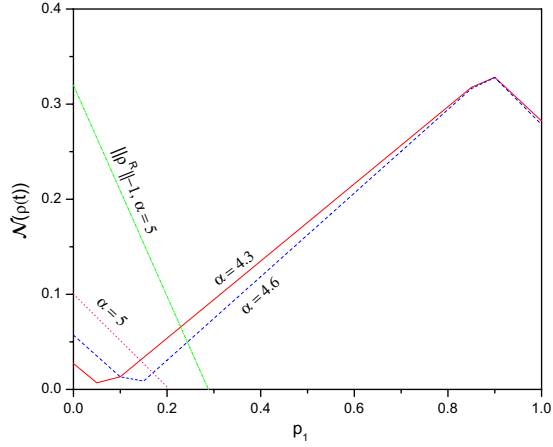}\put(-320,220) &  & 
\end{tabular}%
\end{center}
\caption{(color online) The negativity for various values of parameter $%
\protect\alpha $ is plotted against the decoherence parameter $p_{1}$, for
the case when only one qutrit undergoes decoherence. The realignment
criterion $||\protect\rho ^{R}||-1$ for single qutrit decoherence is also
plotted for $\protect\alpha =5$.}
\label{Figure1}
\end{figure}
The negativity for different values of the parameter $\alpha $, when only
one qutrit is coupled to the environment, is plotted against $p_{1}$ in Fig. 
$1$. It can be seen that the negativity becomes zero at a finite value of $%
p_{1}$ (around $p_{1}=0.2$) only for density matrices that correspond to the
upper limit of the parameter $\alpha $ and hence becomes PPT that might
possess some degree of entanglement. For density matrices in the lower limit
of $\alpha $, the negativity first decreases to a minimum value, however,
still positive and then increases with $p_{1}$, thereby increasing the
degree of NPT (free entanglement). The partial transpose criterion fails to
detect the behavior of entanglement in density matrices of large $\alpha $
beyond certain values of $p_{1}$. Further investigation of the entanglement
in these states can be carried out by using realignment criterion. The
quantity $\left\Vert \rho ^{R}\right\Vert -1$ for density matrices of large $%
\alpha $ is also plotted in Fig. $1$. The positive value of this quantity in
the range $0.21\leq p_{1}\leq 0.288$ shows that these density matrices are
bound entangled. However for $p_{1}>0.288$, the realignment criterion also
fails to detect the possible entanglement for these states. 
\begin{figure}[h]
\begin{center}
\begin{tabular}{ccc}
\vspace{-0.5cm} \includegraphics[scale=1.2]{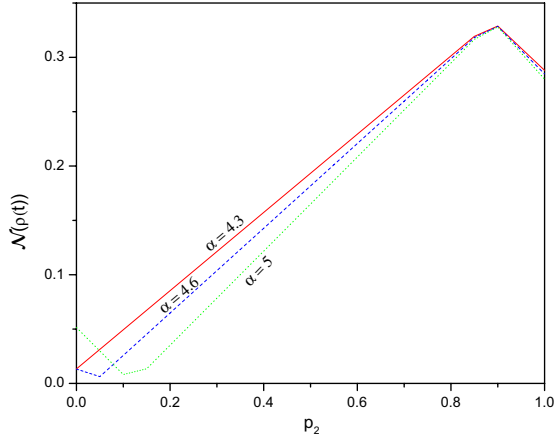}\put(-320,220) &  & 
\end{tabular}%
\end{center}
\caption{(color online) The negativity for various values of parameter $%
\protect\alpha $ is plotted against the decoherence parameter $p_{2}$ when
both the qutrits are coupled to their local environments. The value of
decoherence parameter $p_{1}=0.1$.}
\label{Figure 2}
\end{figure}

The negativity for the case when both qutrits are individually coupled to
its own environment, that is, for multilocal coupling is plotted in Fig. $2$
against the decoherence parameter $p_{2}$ for decoherence parameter $%
p_{1}=0.1$. It can be seen that neither DSD nor ESD occurs in any density
matrix ($4<\alpha \leq 5$) as the negativity is positive for all of them.
For density matrices of lower values of $\alpha $ in the range $4<\alpha
\leq 5$, the negativity increases linearly with $p_{2}$ and hence the degree
of entanglement. On the other hand, for density matrices that correspond to
intermediate and upper values of $\alpha $, it first decreases, however
remain positive, and then increases linearly. It can be shown that the
negativity increases linearly for all density matrices, irrespective of the
value of $\alpha $ ($4<\alpha \leq 5$), for larger values of the decoherence
parameter $p_{1}$. Thus, both DSD and ESD can be completely avoided in the
case of multilocal coupling of the system and environment by controlling the
environment of one qutrit at least. Nevertheless, for $p_{1}<0.1$, some
density matrices that correspond to large values of $\alpha $ become PPT.
For example, for $p_{1}=0.05$, the negativity of density matrix that
corresponds to $\alpha =5$ becomes zero at $p_{2}=0.165$. The negativity and
the realignment criterion for such density matrix is plotted in Fig. $3$. It
shows that the density matrix becomes bound entangled for $0.165<p_{2}\leq 1$%
. 
\begin{figure}[h]
\begin{center}
\begin{tabular}{ccc}
\vspace{-0.5cm} \includegraphics[scale=1.2]{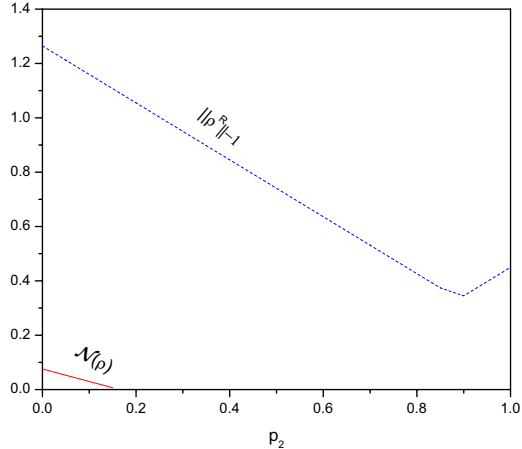}\put(-320,220) &  & 
\end{tabular}%
\end{center}
\caption{(color online) The negativity and realignment criterion $||\protect%
\rho ^{R}||-1$ for the multilocal coupling are plotted against the
decoherence parameter $p_{2}$. The values of other parameters are set to $%
p_{1}=0.05$, $\protect\alpha =5$.}
\label{Figure 3}
\end{figure}
\begin{figure}[h]
\begin{center}
\begin{tabular}{ccc}
\vspace{-0.5cm} \includegraphics[scale=1.2]{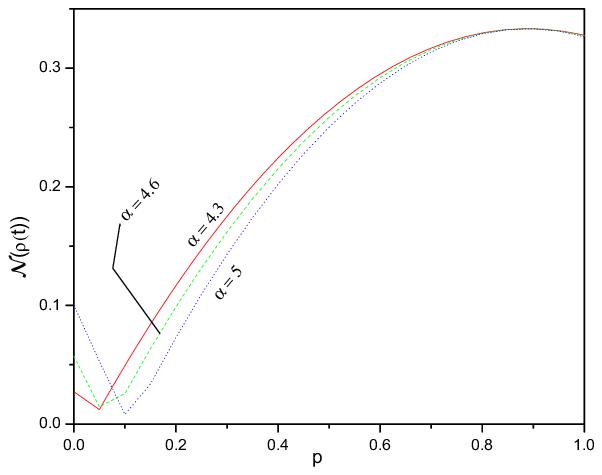}\put(-320,220) &  & 
\end{tabular}%
\end{center}
\caption{(color online) The negativity for various values of $\protect\alpha 
$ is plotted against decoherence parameter $p$ when the system evolves under
collective depolarizing noise.}
\label{Figuer 4}
\end{figure}

In Fig. $4$, we have plotted the negativity for various density matrices
against the decoherence parameter $p$ for the case of collective noise. As
can be seen from the figure, non of the density matrices undergoes DSD or
ESD. According to partial transpose criterion, all the density matrices in
the range $4<\alpha \leq 5$ are free entangled under the action of
collective depolarizing noise. Though the negativity for each density matrix
drops to a minimum, but positive value, in the range of lower values of
decoherence parameter. The degree of NPT increases for all density matrices
with the values of decoherence parameter $p$ in the large values limit. 
\begin{figure}[h]
\begin{center}
\begin{tabular}{ccc}
\vspace{-0.5cm} \includegraphics[scale=1.2]{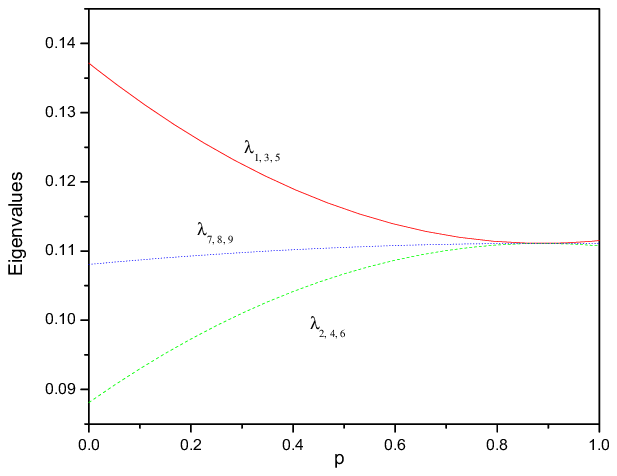}\put(-320,220) &  & 
\end{tabular}%
\end{center}
\caption{(color online) The eigenvalues for the case of global noise are
plotted against the decoherence parameter $p$. The values of other
parameters are set to $p_{1}=p_{2}=0.5$, $\protect\alpha =4.3$.}
\label{Figure 5}
\end{figure}

The general form of eigenvalues for the global noise have quite lengthy
expressions, instead of writing their expression, we prefer to see their
behavior by plotting them against decoherence parameter $p$. Such a plot for 
$p_{1}=p_{2}=0.5$ and $\alpha =4.3$ is shown in Fig. $5$. Since all the
eigenvalues for the whole range of decoherence parameter are positive, the
density matrices under global noise are PPT. The realignment criterion also
fails to detect any possible entanglement as the quantity $\left\Vert \rho
^{R}\right\Vert -1$ for the range of interest of the parameter $\alpha $
against the decoherence parameter $p$ is negative. We conclude that the
distillable free entangled qutrit-qutrit states becomes completely
nondistillable in the presence of global depolarizing noise. Whereas under
the action of global dephasing noise, it is shown \cite{Mazher2} that these
density matrices are free entangled in certain limited time and undergo DSD
at a later time.

To see the effect of local unitary operation on the dynamics of the state of
Eq. \ref{4}, we consider the unitary operator $U=I_{3}\otimes \theta $, with 
$\theta =|0\rangle \langle 1|+|1\rangle \langle 0|+|2\rangle \langle 2|$.
When this operator is applied locally to the state of Eq. \ref{4}, it
converts the maximally entangled state $|\Psi _{+}\rangle $ into another
maximally entangled state given by $|\tilde{\Psi}_{+}\rangle =1/\sqrt{3}%
\left( |00\rangle +|11\rangle +|22\rangle \right) $ and the two separable
states $\sigma _{+}$, $\sigma _{-}$ into other two separable states,
respectively, given by $\tilde{\sigma}_{+}=1/3\left( |00\rangle \langle
00|+|12\rangle \langle 12|+|21\rangle \langle 21|\right) $, $\tilde{\sigma}%
_{-}=1/3\left( |11\rangle \langle 11|+|20\rangle \langle 20|+|02\rangle
\langle 02|\right) $. The density matrix after $U$ acts on $\rho (0)$ becomes%
\begin{equation}
\sigma _{\theta }=U\rho (0)U^{\dagger }=\frac{2}{7}|\tilde{\Psi}_{+}\rangle
\langle \tilde{\Psi}_{+}|+\frac{\alpha }{7}\tilde{\sigma}_{+}+\frac{5-\alpha 
}{7}\tilde{\sigma}_{-}  \label{8}
\end{equation}%
The evolution of $\sigma _{\theta }$ in the noisy environment and its
partial transpose can be found straightforwardly in the same manner as done
for the state $\rho (t)$. It is found that all the eigenvalues for each
individual case, considered for the coupling of $\rho (t)$, remains
unchanged.\ In conclusion, the local operation that we consider here does
not change the behavior of entanglement in the presence of any of the
aforementioned coupling under the action of depolarizing noise. It is
important to state that under the action of amplitude damping noise, a local
unitary operation changes the behavior of entanglement and completely avoids
DSD \cite{Mazher1}.

\section{Summary}

We study the dynamics of entanglement for particular bipartite qutrit
density matrices under global, collective, multilocal and local depolarizing
noise. We show that unlike the cases of dephasing \cite{Mazher2} and
amplitude damping noises \cite{Mazher1}, the influence of depolarizing noise
is completely different. Using partial transpose criterion and realignment
criterion, it is shown that only those density matrices that correspond to
large values of $\alpha $ becomes PPT and thus nondistillable in the local
depolarizing noise. In the case of multilocal depolarizing noise, both ESD
and DSD can be avoided by controlling one or the other local environment. We
also show that instead of ESD and DSD to occur, the degree of entanglement
for free entangled density matrices under collective depolarizing noise
increases with the increasing value of decoherence parameter. All the free
entangled density matrices in the specified range of parameter $\alpha $
becomes PPT under the action of global depolarizing noise. In conclusion, we
show that the free entangled distillable density matrices convert into bound
entangled or separable density matrices and thus become completely
nondistillable under global noise. Also, the depolarizing noise can be used
to increase the degree of entanglement in the specific family of density
matrices that correspond to a particular values of the parameter $\alpha $
under certain couplings with the environment. Furthermore, It is shown that
the local operation, we used in this paper, does not change the dynamics of
entanglement under each coupling of the system and environment for the case
of the specific family of the density matrices. Since there is no single
known definitive criterion for separability and entanglement in states of
dimension greater than six, therefore, we emphasis that the validity of our
results is true for the family of density matrices that we consider here and
may not be generalized to all bipartite qutrit states.\newline

{\LARGE Figure Captions}\newline
Figure $1$. The negativity for various values of parameter $\alpha $ is
plotted against the decoherence parameter $p_{1}$, for the case when only
one qutrit undergoes decoherence. The realignment criterion $||\rho ^{R}||-1$
for single qutrit decoherence is also plotted for $\alpha =5$.\newline
Figure $2$. The negativity for various values of parameter $\alpha $ is
plotted against the decoherence parameter $p_{2}$ when both the qutrits are
coupled to their local environments. In the inset of the figure the
realignment criterion is plotted against decoherence parameter $p_{2}$,
which shows that the density matrices become bound entangled for $p_{N}=$ $%
0.165<p_{2}\leq 1$.\newline
Figure $3$. The negativity and realignment criterion $||\rho ^{R}||-1$ for
the multilocal coupling are plotted against the decoherence parameter $p_{2}$%
. The values of other parameters are set to $p_{1}=0.05$, $\alpha =5$.%
\newline
Figure $4$. The negativity for various values of $\alpha $ is plotted
against decoherence parameter $p$ when the system evolves under collective
depolarizing noise.\newline
Figure $5$. The eigenvalues for the case global noise are plotted against
the decoherence parameter $p$. The values of other parameters are set to $%
p_{1}=p_{2}=0.5$, $\alpha =4.3$.

\end{document}